\documentclass{ifacconf}
\usepackage{natbib}        

\usepackage{amsmath}
\usepackage{graphicx}      
\usepackage{cases}

\usepackage{graphicx}
\usepackage{url}

\usepackage{xspace}
\usepackage{amsfonts}

\newcommand{\stam}[1]{}
\newcommand{\comment}[1]{}

\newcommand{\Yhankel}{Y_{0,r,N}\,}
\newcommand{\Xhankel}{X_{0,1,N}\,}
\newcommand{\Uhankel}{U_{0,r,N}\,}
\newcommand{\Pihankel}{\Pi_{0,r,N}\,}
\newcommand{\RO}{\text{range}(O_r)}
\newcommand{\vy}{{\bf y}}
\newcommand{\ve}{{\bf e}}
\newcommand{\ym}{y_{m}}
\newcommand{\vym}{{\bf y_{meas}}}
\newcommand{\lone}{\lambda_1}
\newcommand{\ltwo}{\lambda_2}
\newcommand{\lonemax}{\lambda_1^{max}}
\newcommand{\ltwomax}{\lambda_2^{max}}

\DeclareMathOperator*{\argmin}{arg\,min}

 \def\B{{ \bf B}}
\def\V{{ \bf V}}

\def\Y{{\boldsymbol{Y}}}

\def\subjto{{\mbox{subj. to}}}

\newcommand{\T}{\mathsf{T}}

\newtheorem{rmk}{Remark}

\begin{document}
\begin{frontmatter}

\title{Robust Subspace System Identification via Weighted Nuclear Norm Optimization\thanksref{footnoteinfo}} 

\thanks[footnoteinfo]{Ohlsson gratefully acknowledge support from
  the NSF project FORCES (Foundations Of Resilient CybEr-physical Systems), the Swedish Research
  Council in the Linnaeus center CADICS, the European Research Council
   under the advanced grant LEARN, contract 267381, a postdoctoral grant from the Sweden-America
   Foundation, donated by ASEA's Fellowship Fund, and  by a postdoctoral
   grant from the Swedish Research Council.}

\author[All]{Dorsa Sadigh} 
\author[All,Liu]{Henrik Ohlsson} 
\author[All]{S. Shankar Sastry}
\author[All]{Sanjit A. Seshia}

\address[All]{University of California, Berkeley, 
Berkeley, CA 94720 USA. {\tt\{dsadigh,ohlsson,sastry,sseshia\}@eecs.berkeley.edu}}

\address[Liu]{Division of Automatic Control, Department of Electrical Engineering, Link\"oping University, Sweden.}
\begin{abstract}                
Subspace identification is a classical and very well studied problem
in system identification. The problem was recently posed as a convex
optimization problem via the nuclear norm relaxation. Inspired by
robust PCA, we extend this framework to handle
outliers. The proposed framework takes the form of a convex
optimization problem with an objective that trades off fit, rank and
sparsity.  As in robust PCA,  it can be problematic to find a suitable
regularization parameter. We show how the space in which a suitable
parameter should be sought can be limited to a bounded open set of the
two-dimensional parameter space. In practice, this is very useful
since it restricts the parameter space that is needed to be surveyed.
\end{abstract}


\end{frontmatter}

\section{Introduction}
\label{sec:intro}

\label{sec:background}
%
%
Subspace system identification is a well studied problem within the field of system identification~\citep{DeMoor1998,Moonen89,verhaegen1992subspace,verhaegen1992subspaceb,verhaegen1993subspacec,Verhaegen1994,VanOverschee1994}. The problem has a nice geometrical interpretation and can be posed as a rank minimization problem. However, minimizing the rank of a matrix is NP-Hard. 
In recent years, there has been an increasing interest in applying the nuclear norm as a relaxation of  the rank~\citep{fazel2013,liu2010,liu2013,hansson2012}. The nuclear norm, which is the sum of singular values, gives a convex approximation for the rank of a matrix. Thus, it provides a convenient framework for system identification as well as preserving the linear structure of the matrix. In this approach, the nuclear norm of a Hankel matrix representing the data and a regularized least square fitting error is minimized. \citet{fazel2013} study the problem of rank minimization and compare the time complexity of different algorithms minimizing the dual or primal formulation.

The problem of subspace identification with partially missing data is addressed by~\citet{liu2013}, where they extend a subspace system identification problem to the scenario where there are missing inputs and outputs in the training data. The authors approach this case by solving a regularized nuclear norm optimization, where the least square fitting error is only minimized over the observed data.

Low rank problems have 
also been studied extensively in the areas of machine learning and statistics. One of the most studied problems is that of principal component analysis (PCA, \citet{Hotelling:33}). Although both of the  approaches, the subspace identification framework and  PCA, seek low-rank structures, the major difference is the additional structure imposed in the subspace identification framework due to the linear dynamics of the system. 

In this work, we extend the nuclear norm minimization framework to the case, where the output data has outliers. Our framework considers a situation where the observed sensors are attacked by a malicious agent. Thus, we would like our subspace system identification approach to be resilient to such attacks. In our solution, we formalize three tasks: (i) detecting the attack vector,  (ii) minimizing the least square fitting error between our estimation and the training data, (iii) rank minimization. 
The attack vector is assumed to be sparse with nonzero entries corresponding to the instant of attack.
We do not impose any structure on the time of attack, which is the position of outliers in the attack vector. We then estimate the attack vector as well as the model orders and model matrices. 
In order to impose the trade off between sparsity of the attack vector and simplicity of the structure of the system, both the attack term and the nuclear norm are penalized.

Our approach is inspired by the developed techniques in machine learning and robust PCA. The problem of robust PCA \citep{candes2011}, that is to find the principal components when there exists corrupted
training data points or outliers is of interest in applications like image reconstruction. The common solutions of this problem include using a robust estimator for covariance matrix.

The main contributions of the paper are twofold. The first contribution is a novel framework for robust subspace identification. The method is based on convex optimization and accurately detects outliers. The second contribution is the characterization of the regularization parameter space that needs to be surveyed. More precisely, we  show that the optimization variables are zero outside a bounded  open set of the two-dimensional parameter space and that the search for suitable regularization parameters can therefore be limited to this set.
The derivations also apply after minor modifications to limit the search space for algorithms for robust PCA \citep{candes2011} and subspace identification \citep{fazel2013,liu2010,liu2013,hansson2012}.

In the rest of this paper, we first propose our problem setting in Section~\ref{sec:probdef}. We then discuss our method for detecting outliers in Section~\ref{sec:method}, and propose a heuristic for computing the penalty terms that we introduce in Section~\ref{sec:penalty}. In Section~\ref{sec:expts} we implement our algorithm and show the results for a dataset. We then conclude in Section~\ref{sec:conclusion}.

\section{Problem Formulation}
\label{sec:probdef}
The problem of subspace identification can be formulated for a linear
discrete-time state space model with process and measurement noise.
We use the following Kalman normal form for this formulation.

\begin{equation} \label{state-space}
\begin{array}{ll}
x(k+1) = Ax(k) + Bu(k) + Ke(k)\\
y(k) = Cx(k) + Du(k) + e(k)
\end{array}
\end{equation}

In equation~\eqref{state-space}, we let $x(k) \in {\bf R}^{n_x}, u(k)
\in {\bf R}^{n_m}, y(k) \in {\bf R}^{n_p}$ and $e(k) \in {\bf R}^{n_p}$, where $u(k)$ is the set of inputs, and $y(k)$ is the set of outputs. We let $e(k)$ be ergodic, zero-mean, white noise. Matrices $A, B, C, D, K$ are real valued system matrices of this state-space model. The problem of subspace identification is to estimate system matrices and model order $n_x$, given a set of input and output traces $(u(k), y(k))$ for $k = 0, \cdots, N$.

In this work, we consider a variant of subspace identification problem, where we experience missing data and outliers in the set of output traces. Our goal is to estimate the system matrices and model orders correctly in the presence of such outliers and missing data.

Throughout this paper, we use block Hankel matrix formulation as in \citep{hansson2012,liu2013} to represent equation~\eqref{state-space}.

\begin{equation}\label{hankel}
\Yhankel = O_r\, \Xhankel + S_r\, \Uhankel + E
\end{equation}

Here, $\Xhankel, \Yhankel$ and $\Uhankel$ are block Hankel matrices for the state, output and input sequences. A block Hankel matrix $H_{i,j,k}$ for a sequence of vectors $h(t)$ is defined to be:

\begin{equation*}
H_{i,j,k} = \begin{bmatrix}
h(i) & h(i+1) & \cdots &h(i+k-1)\\
h(i+1) & h(i+2) & \cdots &h(i+k)\\
\vdots & \vdots &\ddots & \vdots \\
h(i+j-1) & h(i+j) & \cdots &h(i+j+k-2)
\end{bmatrix}
\end{equation*}

In equation~\eqref{hankel}, $E$ is the noise sequence contribution, and $O_r$ is the extended observability matrix. $O_r$ and $S_r$ are defined as the following matrices:

\begin{equation*}
O_r = \begin{bmatrix}
C\\
CA\\
CA^2\\
\vdots\\
CA^{r-1}
\end{bmatrix}, \quad
S_r = \begin{bmatrix}
D & 0 & \cdots & 0 \\
CB & D & \cdots & 0\\
CAB & CB & \cdots & 0\\
\vdots & \vdots & \ddots & \vdots \\
CA^{r-2}B & CA^{r-3}B & \cdots & D
\end{bmatrix}
\end{equation*}

The approach introduced by~\citet{liu2013}, estimates the range space of $O_r$, which then can be used to attain the system matrices. 
First, the second term in equation~\eqref{hankel} is eliminated by multiplying both sides of the equation by $\Pihankel$, an orthogonal projection matrix onto the nullspace of $\Uhankel$.

\begin{equation}\label{pi}
\Yhankel \Pihankel = O_r\, \Xhankel \Pihankel + E\, \Pihankel
\end{equation}

As a result, in the absence of noise the following equality holds:

\begin{equation}
\text{range}(\Yhankel \Pihankel) = \RO
\end{equation}

If $\Xhankel \Pihankel$ has full rank (which is generally the case for
random inputs), it can be shown that \linebreak $\text{rank}(\Yhankel \Pihankel)$ is equal to $n_x$. Therefore, $\RO$ and consequently the model order $n_x$ can be determined by low-rank approximation of $\Yhankel\Pihankel$. 

In order to guarantee the convergence of  \linebreak $\text{range}(\Yhankel \Pihankel)$ to $\RO$ as the number of input, output sequences approach infinity, a matrix $\Phi$ consisting of instrumental variables is introduced.

\begin{equation}
\Phi = \begin{bmatrix} U_{-s,s,N}\\Y_{-s,s,N}\end{bmatrix}
\end{equation} 

We choose $s$ and $r$ to be smaller than $N$. To further improve the accuracy of this method, we include weight matrices $W_1$ and $W_2$. Therefore, the problem of low-rank approximation of $\Yhankel \Pihankel$ is reformulated as low-rank approximation of $G$.

\begin{equation}
\label{eq:G(y)}
G = W_1\,\Yhankel\Pihankel \Phi^\top\, W_2
\end{equation}

The weight matrices that are selected in our experiments are $W_1 = I$, and $W_2 = (\Phi\,\Pihankel\,\Phi^\top)^{-1/2}$ as used in PO-MOESP algorithm by~\citet{Verhaegen1994}.

We approximate $\RO$ to be $\text{range}(W_1^{-1}\,P)$, where $P$ is extracted from truncating the SVD of $G$:

\begin{equation}
G = 
\begin{bmatrix}
P & P_e
\end{bmatrix}
\begin{bmatrix}
\Sigma & 0 \\
0 & \Sigma_e
\end{bmatrix}
\begin{bmatrix}
Q & Q_e
\end{bmatrix}^\top
\end{equation}

After estimation of $\RO$, we can find the matrix realization of the system, and completely recover $A, B, C, D$ and $x_0$.

We let $V \in {\bf R}^{rn_p \times n_x}$ be a matrix whose columns are a basis for the estimate of $\RO$. Then we partition $V$ into $r$ block rows $V_0, \cdots , V_{r-1}$. Each of these blocks has size of $n_p \times n_x$. Then the estimates of $A$ and $C$ are:

\begin{equation}
\label{eq:AC}
\hat{C} = V_0, \quad \quad \hat{A} = \argmin \sum_{i=1}^{r-1} \| V_i \, -\, V_{i-1} \hat{A} \|_F^2
\end{equation}

In this equation $\|\cdot \|_F$ is the Frobenius norm. Based on the estimates $\hat{A}, \hat{C}$ it is easy to solve the following optimization problem that finds $\hat{B}, \hat{D}$ and $\hat{x}_0$.

\begin{equation}
\label{eq:BD}
\begin{array}{ll}
(\hat{B}, \hat{D}, \hat{x}_0)= \argmin \sum_{k=0}^{N+r-2} \| \hat{C} \hat{A}^k \hat{x}_0 + \\
\sum_{i=0}^{k-1} \hat{C} \hat{A}^{k-i} \hat{B} u(i) + \hat{D} u(k) - y(k) \|_2^2
\end{array}
\end{equation}

\section {Method}
\label{sec:method}
\citet{liu2010} approach the subspace identification problem with missing data using a nuclear norm optimization technique. A nuclear norm of a matrix $||X||_*  = \sum_i \sigma_i(X)$ is the sum of all singular values of matrix $X$, and it is the largest convex lower bound for rank($X$) as shown by~\citet{Fazel2001}. 

Therefore, given a sequence of input and output measurements, \citet{liu2013} formulate the following regularized nuclear norm problem to estimate $\vy = y(0), \cdots, y(N+r-2)$, which is a vector of model outputs.

\begin{equation}
\label{eq:nucnorm}
\min_{\vy} \|G(\vy)\|_* + \lambda \, \sum_{k \in T_o} \| y(k) - y_{meas}(k) \|_2^2
\end{equation}

In equation~\eqref{eq:nucnorm}, $T_o$ is the set of time instances for observed output sequences and $T_o \subseteq T$, where $T=\{0,\cdots, N+r-2\}$. Here $y_{meas}(k)$ are the set of measured outputs and $k \in T_o$.
 The first element of the objective function is the nuclear norm of $G(\vy) = W_1\,\Yhankel\Pihankel \Phi^\top\, W_2$, as it is derived in equation~\eqref{eq:G(y)}.

%

Our approach, in detecting output outliers in the training data is built based on the introduced technique. We assume that the vector ${\bf y_{meas}}$, the measured output vector, has a sparse number of outliers or attacked output values. We do not make any extra assumptions on the specific time that the outliers will occur. Thus, we can extend equation~\eqref{eq:nucnorm} by introducing an error term $e(k) \in {\bf R}^{n_p}$ for $k \in T$. This error term is intended to represent the outlier present at time $k$; therefore, we would like vector ${\bf e}$ to be sparse and its non-zero elements detect the time and value of the outliers. Then, the objective we would like to minimize is:

\begin{equation}
\label{eq:outlier}
\begin{array}{lll}
\min_{\vy,\ve} \; f(\vy, \ve; \lone, \ltwo)  = & &\\
\min_{\vy, \ve} \;  \lone \|G(\vy)\|_* + \sum_{k \in T}   \| y(k) - y_{meas}(k) -e(k)\|_2^2  & \\
\quad \quad + \ltwo \sum_{k \in T} \|e(k)\|_1&
\end{array}
\end{equation} 

In this formulation, we would like to estimate $\vy$ and find the error term $\ve$, such that the error vector is kept sparse, and accounts for the outlier values that occur in training data $\vym$. 
The first term in this formulation is the nuclear norm with a penalty term $\lone$. The second term is the least square error as before, which enforces  $\ve$ to capture the outlier values in ${\bf y_{meas}}$. The last term is the $\ell_1$-norm enforcing the sparsity criterion on vector $\ve$. We penalize the $\ell_1$-norm with $\ltwo$. 

Using the formulation in equation~\eqref{eq:outlier}, allows us to:
\begin{enumerate}
\item find a filtered version of the output measurements. In this filtered version of the output, the effects of outliers have been removed and the values for the missing data are filled in.   
 \item In addition, \eqref{eq:outlier}  allows us to get an estimate of the value and time the outliers are appeared in the measurements. This is a valuable piece of information if the time of attack is a variable of interest.
\end{enumerate}
Having recovered the filtered output, any subspace identification method could be applied to estimate the model matrices $A$, $B$, $C$, $D$ and $K$.

\section{Penalty Computation}
\label{sec:penalty}
In this section, we discuss how to choose the penalty terms $\lone$
and $\ltwo$ in equation~\eqref{eq:outlier}. 
Notice that a large enough $\lone$ will force $\vy=0$ and  a large
enough $\ltwo$ drives $\ve=0$.

In fact, it can be shown that there exist
$\lonemax$ and $\ltwomax$ such that whenever $\lone \geq \lonemax$,
$\vy=0$  and whenever $\ltwo \geq \ltwomax$, $\ve=0$.  In practice,
$\lonemax$ and $\ltwomax$ are very useful since they give a range for
which it is interesting to seek good penalty values. 
Having limited the search for good penalty values to an open set in
the $(\lone,\ltwo)$-space, classical model selection techniques such
as cross validation or
 the Akaike criterion (AIC) \citep{Akaike1973} could be adapted to
find suitable penalty values. 

\subsection{Computation of $\lonemax$ and $\ltwomax$}
\label{lambda_max}
The optimal solution of equation~\eqref{eq:outlier} occur only when zero is included in the subdifferential of the objective in equation~\eqref{eq:outlier}.
\begin{equation}
0 \in \partial f(\vy,\ve; \lone, \ltwo)
\end{equation}

We find the values $\lonemax$ and $\ltwomax$, by solving $0 \in \partial f(\vy,\ve; \lone, \ltwo)$ subject to the constraints $\vy = 0$ and $\ve = 0$. 

For simplicity, assume  that $n_p=1$. Therefore, equation~\eqref{eq:outlier} can be simplified:

\begin{equation}
\label{outlier2}
\min_{\vy,\ve} \; \ltwo\|G(\vy)\|_* + \sum_{k \in T}   ( y(k) -
\ym (k) - e(k))^2 + \lone |e(k)|
\end{equation} 

This equation can be reformulated using the Huber norm \citep{Huber1973}:
\begin{equation}
\label{outlier3}
\min_{\vy} \; \ltwo\|G(\vy)\|_* + \sum_{k \in T}   \| y(k) - \ym (k) \|_H,
\end{equation}
where the Huber norm $\|\cdot  \|_H $ is defined:

\begin{equation}
 \| x \|_H=\begin{cases} x^2 & \text{if } |x|\leq \lone/2, \\   \lone
  | x|-\lone^2/4 &\text{otherwise} \end{cases}
\end{equation}

Note that $\|\cdot  \|_H $ is differentiable. Now, to find
$\lonemax$ and $\ltwomax$ we seek the
smallest $\ltwo$ such that $0$ belongs to the subdifferential of the
objective function with respect to $\vy$ evaluated at $\vy = 0$.

\begin{equation}
\label{eq:zero}
0\in \partial_{\vy}  \Big( \ltwo\|G(\vy)\|_*
+ \sum_{k \in T}   \| y(k) - \ym (k) \|_H \Big) \Big|_{\vy=0}
\end{equation}

This subdifferential with respect to $y(t)$ can be calculated: 

\begin{subequations}
\label{eq:partial}
\begin{align}
 \partial_{\vy(t)} & \Big( \ltwo \|G(\vy)\|_*
+ \sum_{k \in T}   \| y(k) - \ym (k) \|_H \Big) \\
= &  \ltwo \partial_{\vy(t)}  \Big(\|G(\vy)\|_* \Big)
+\partial_{\vy(t)}  \Big(   \| y(t) - \ym (t)
\|_H \Big) 
\end{align}
\end{subequations}

In equation~\eqref{eq:G(y)}, we defined $G(\vy)$. Since we chose $W_1$ to be the identity matrix $I$, it is reasonable to assume that $G(\vy)$ takes the form:

\begin{equation}
G(\vy) = \Yhankel \B 
\end{equation}

Thus, the subdifferential of the nuclear norm in equation~\eqref{eq:partial} evaluated at $\vy=0$ is:

\begin{equation}
\label{eq:first}
\begin{array}{lll}
 \partial_{\vy(t)}  \Big(\|G(\vy)\|_* \Big) \Big |_{\vy=0}&=&    \ltwo \sum_{i,j} \V(i,j) 
\partial_{\vy(t)} \Big ( G(\vy)(i,j) \Big) \Big |_{\vy=0}  \\
&=&\sum_{k=1}^t
\V(t-k+1,:) \B(k,:)^\T 
\vspace{3mm} 
\\
&& \text{where}\; \|\V\|\leq 1
\end{array}
\end{equation}

See \citet{Watson199233} and \citet{recht10} for the calculation of
the subdifferential of the nuclear norm.

Equation~\eqref{eq:first} is analyzed for $t=1,\dots, N+r-1$. We calculate this subdifferential separately for three different intervals of $t$: (i) $t=1,\dots, r$, (ii) $t=r+1,\dots, N$, (iii) $t=N+1,\dots, N+r-1$ due to the structure of the block Hankel matrix $Y_{0,r,N}$. 

Furthermore, we calculate the subdifferential of the second part of equation~\eqref{eq:partial}:

\begin{equation}
\label{eq:second}
\begin{array}{l}
\partial_{\vy(t)}  \Big(   \| y(t) - \ym (t)
\|_H \Big)= 
\vspace{1mm} 
\\
\begin{cases} 2\big (y(t) - \ym (t) \big) & \text{if } |y(t) - \ym (t)|\leq \lone/2, \\   \lone
   \text{sgn} \big( y(t) - \ym (t) \big)  &\text{otherwise}\end{cases}
 \end{array}
\end{equation}

Combining the two parts in equations~\eqref{eq:first} and \eqref{eq:second},  we rewrite the subdifferential of the objective function.

 \begin{equation}
 \begin{array}{l}
 0\in \ltwo\sum_{k=1}^t
\V(t-k+1,:) \B(k,:)^\T
-\vspace{2mm} 
\\
\begin{cases} 2 \ym (t) & \text{if } |\ym (t)|\leq \lone/2\\   \lone
   \text{sgn} \big (\ym (t) \big)  &\text{otherwise} \end{cases},
  \quad  \|\V\|\leq 1
\end{array}
\end{equation}

We can hence find $\ltwomax$ (for
each value of $\lone$) by solving the following convex program: 

\begin{equation}
\label{eq:l2max}
\begin{array}{llll}
\ltwomax &=& \argmin_{\vy} \|\V\| &
\vspace{2mm} 
 \\
&& \subjto & 0=\sum_{k} \; \V(t-k+1,:) \B(k,:)^\T - \\
&& &\begin{cases} 2 \ym (t) & \text{if } |\ym (t)| \leq \lone/2\\   
\lone \text{sgn} \big (\ym (t) \big)   &\text{otherwise} \end{cases}
\vspace{2mm} 
 \\
&&\text{for  } t=1,\dots, r. &
\end{array}
\end{equation}

\begin{rmk}
\label{remark}
Note that if $\lambda_1$ is chosen such that $ |\ym (t)| \leq
\lone/2,$  for  $t=1,\dots,N+r-1$, then $\ve=0$ solves
\eqref{eq:outlier}. Therefore, $\lonemax=2 \max_t |\ym (t)|$.
\end{rmk}

\begin{rmk}
\label{remark2}
In all our calculations, $\ltwomax$ is a function of $\lone$. From now on, we refer to $\ltwomax$ as the value of this function evaluated at $\lone = \lonemax$.
\end{rmk}

Therefore, based on remarks~\eqref{remark} and \eqref{remark2} and by solving equation~\eqref{eq:l2max} we find $\lonemax$ and $\ltwomax$ for a given sequence of inputs and outputs
\subsection{Finding the knee of the residual curve}
For a given $\lone$ and $\ltwo$, the residual training error is the sum of squared residual errors at every time step of the training data, which is the difference between the measured $\ym(t)$ and the simulated $\tilde{y}(t)$ and the error term $e(t)$. The simulated $\tilde{y}(t)$ is the output of simulation of the dynamics after estimating the system matrices based on equations~\eqref{eq:AC} and~\eqref{eq:BD}. The term $e(t)$ encodes the position and amount of outliers that occur in the training measured data $\ym(t)$.

\begin{equation}
\label{residual}
\text{Residual Training Error} = \sum_{t \in T} (\tilde{y}(t) + e(t) - \ym(t))^2
\end{equation}

\begin{figure}[h]
\centering
\includegraphics[width=0.5\textwidth]{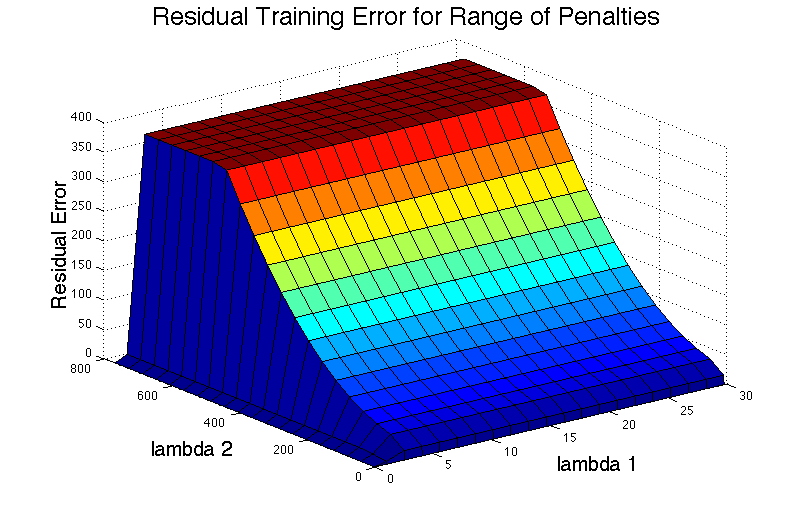}
\caption{\small Plot of the residual training error for combination of $(\lone, \ltwo)$. The residual training error increases as both $\lone$ and $\ltwo$ are increasing.}
\label{fig:surface}
\end{figure}

 Since we have found $\lonemax$ and $\ltwomax$, we are now able to grid over the intervals $(0,\lonemax]$ and $(0,\ltwomax]$, and calculate the residual training error for every point in the grid. Figure~\ref{fig:surface} represents this residual error for combinations of $(\lone,\ltwo) \in (0,\lonemax] \times (0,\ltwomax]$. We linearly grid each one of the intervals the two penalty terms lie in. In this example given $\lonemax = 29.6682$ and $\ltwomax = 766.8142$, we pick 20 linearly spaced values for each $\lone$ and $\ltwo$ as shown in Figure~\ref{fig:surface}.

Given a fixed $\lone$, the graph of the residual training error increases as $\ltwo$ increases. We would like to pick a value for $\ltwo$ that minimizes this error; however, we must avoid overfitting which can be caused by picking the smallest possible $\ltwo$. Therefore, we select $\ltwo$ such that it is at the knee of the curve.
The knee of a curve is the point on the curve where the rate of performance gain starts diminishing, which is the area with the maximum curvature. We select the most effective $\ltwo$ (or $\lone$) by choosing the knee of the residual training error curve for a fixed $\lone$ (or $\ltwo$).

\subsection{Cross Validation}
Another approach that could be used to choose $\lone$ and $\ltwo$ is
to perform the calculations in Section~\ref{lambda_max} only on, say, $90\%$ of the data, and calculate the residual error on the validation data which is only the $10\%$ of the dataset that is not used in training. We assume there are no outliers in the validation data; however, subspace system identification cannot be done solely on this batch of data since the $10\%$ validation batch does not have sufficient number of data points for a complete subspace system identification.
After the computation of $\lonemax$ and $\ltwomax$, we choose a grid for combinations of $(\lone, \ltwo)$ as  before. We then calculate the residual validation error for every pair of $(\lone,\ltwo)$. 

\begin{equation}
\label{validation}
\text{Residual Validation Error} = \sum_{t \in T_{valid}} (\tilde{y}(t) -  y_{valid}(t))^2
\end{equation}

The residual validation error is the sum of squared error, where $T_{valid}$ represents the data points for the $10\%$ validation batch, and $y_{valid}(t)$ is the measured output for this batch. As in equation~\eqref{residual}, $\tilde{y}(t)$ is the output of simulation of dynamics for the specific time window $T_{valid}$.
 
In this case, we choose the set of $(\lone,\ltwo)$ that minimize the residual validation error. As both penalty terms get smaller the validation error drops as well; however, this error reaches a minimum and starts increasing as the penalty terms get smaller due to over fitting. Therefore, the most effective set of $(\lone, \ltwo)$ are the largest pair that drive the validation error to its minimum. 

%



\section{Experimental Results}
\label{sec:expts}
\begin{figure}[h]
\centering
\includegraphics[width=0.5\textwidth]{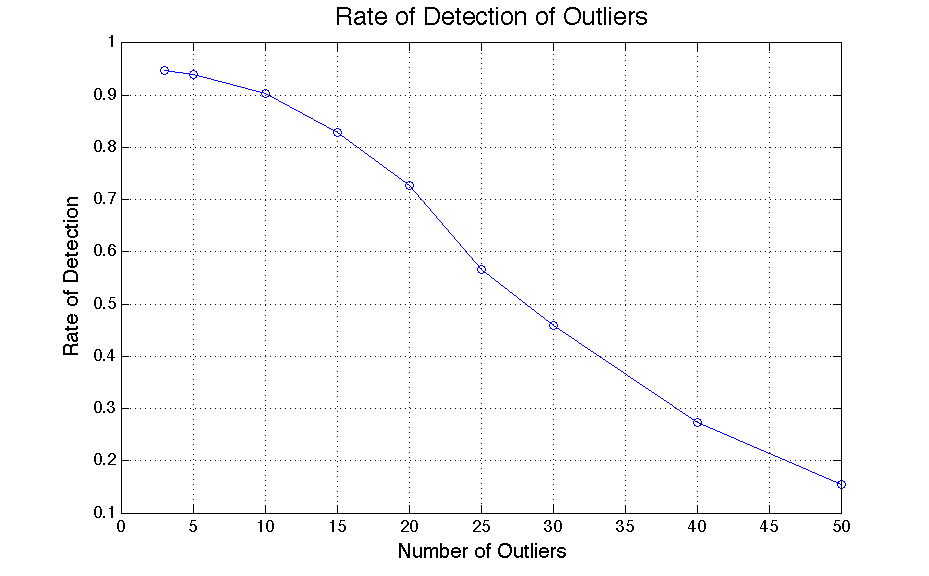}
\caption{\small This plot shows the rate of detection of outliers for a given number of inserted outliers. The rate of detection decreases as we insert more outliers in the data.}
\label{fig:outlier}
\end{figure}

In our experiments we use the data from the DaISy database by~\citet{demoor1997} and insert outliers at randomly generated indices in the output set. We then perform our algorithm to detect the indices with outliers and recover the output $\hat{\vy}$ as well as the model matrices. 
Table~\ref{table:noise} and Figure~\ref{fig:outlier} correspond to data of a simulation of an ethane-ethylene destillation.
We let $r=5$, and $s=5$, and take the first $5$ input and output values of this benchmark as instrumental variables. 
We use the rest of this data sequence as $\vym(t)$, where $t \in \{1,\dots, 85\}$ and $T=85$. 
Then outliers are inserted at randomly chosen indices of $\vym$. 
The insertion of outliers is either by subtracting or adding a large value to a randomly selected index of vector $\vym(k)$.
For the destillation benchmark, $n_p =3$, that is $\vym(k)\in\mathbb{R}^3$. Thus, for a randomly selected time index $k\in\{1,\dots,85\}$, we randomly choose one of the vector elements of $\vym(k)$, and either add or subtract a large value (in our example $20$ since the elements of $\vym$ range from $[-9.5267,7.2139])$.
Based on the calculations  for computation of penalties in Section~\ref{sec:penalty}, we select $\lone=1$ and $\ltwo=1$.

We define rate of correct detection as the ratio of correctly detected outliers to the number of true outliers. The correctly detected outliers are the number of outliers detected at the same exact indices as the true outliers. 
We first set the number of true outliers to $3$ in a dataset with $85$ points, and perform a Monte Carlo simulation with 50 iterations. In every iterations, we insert three randomly chosen outliers in the dataset. We report the mean value of rate of correct detection in Table~\ref{table:noise} for the desillation benchmark. As the numbers in Table~\ref{table:noise} suggest, the rate of correct detection decreases as the noise level of this dataset increases. 

\begin{table}[h]
\label{table:noise}
\centering
\begin{tabular}{| l | c | }
  \hline
  Noise Level & Rate of Outlier Detection  \\ \hline
  \hline                     
  No noise & 0.9800 \\ \hline
  10\% noise & 0.9467 \\ \hline
  20\% noise& 0.8933 \\ \hline
  30\% noise & 0.9000 \\
  \hline  
\end{tabular}
 \caption{Rate of correct outlier detection for ethane-ehylene destillation benchmark with different levels of noise. The penalties are chosen to be $\lone = 1$ and $\ltwo = 1$.}
\end{table}

We then calculate the rate of correct detection for different number of outliers in the dataset. Figure~\ref{fig:outlier} shows the drop in rate of correct detection of outliers as the number of outliers increase in the dataset. Similar to before, we perform 50 iterations of Monte Carlo Simulation and plot the mean value of rate of detection. In every iteration, a new set of randomly generated indices were selected for inserting outliers. We range the number of outlier insertions from $3$ to $50$ points for a dataset with $T=85$ data points. 

We did not encounter any false positives, where the algorithm detects an incorrect index as an outlier point, in any of these iterations. The inexact rate of detection in Figure~\ref{fig:outlier}  corresponds to failure in detection of an outlier in every iteration rather than misdetection. 

With our algorithm, we are able to correctly detect outliers. After detection of the outliers, we can then use any proposed subspace system identification method to find the model matrices. We follow the same approach proposed by~\citet{liu2013}, that is to use the estimation of the output and create the estimated output Hankel matrix $\hat{Y}_{0,r,N}$. Then $\RO$ the extended observability matrix can be evaluated by applying SVD on the estimated $\hat{G}$ as in equation~\eqref{eq:G(y)}.

\begin{equation}
\label{eq:G(y)_est}
\hat{G} = W_1\,\hat{Y}_{0,r,N}\Pihankel \Phi^\top\, W_2
\end{equation}

We then apply the optimizations in equations~\eqref{eq:AC} and~\eqref{eq:BD} to find the estimated model matrices and the estimated initial state. Finally, we have completely realized the system matrices from a dataset with randomly attacked output values.



\section{Conclusion and Future Work}
\label{sec:conclusion}

In this paper, an outlier-robust approach to subspace identification was
proposed. The method takes the form of a convex optimization problem 
and was shown to accurately detect outliers.  The method has two
tuning parameters trading off the sparsity of the estimate for
outliers, the rank of a system matrix (essentially the order of the
system) and the fit to the training data. To aid in the tuning of
these parameters, we show that the search for suitable parameters can
be restricted to  a bounded open set  in the two dimensional parameter
space. This can be very handy in practice since an exhaustive  search
of the two-dimensional parameter space is often time consuming. 
we note that this way of bounding the parameter space could also be
applied to robust PCA etc. This is left as feature work.

To further speed up the framework,  the alternating direction method
of multipliers (ADMM) could be used to solve the optimization
problem. This was not considered here but is seen as a possible
direction for future work.

\bibliography{refs}

\end{document}